\documentclass[sigconf]{acmart}

\usepackage{float}
\usepackage{color}
\usepackage{enumitem}
\usepackage[T1]{fontenc} 
\usepackage{graphicx}
\usepackage{microtype} 
\usepackage{multirow}
\usepackage{rotating}
\usepackage{subcaption}
\usepackage{textcomp}
\usepackage[table]{xcolor}
\usepackage{url}
\usepackage{siunitx}
\usepackage{tikz} 

\definecolor{RoyalBlue}{rgb}{0.255,0.410,0.879}
\definecolor{DarkGreen}{rgb}{0.0,0.5,0.0}
\definecolor{Maroon}{rgb}{0.5,0.0,0.0}
\definecolor{RoyalPurple}{rgb}{0.4,0.0,0.23}
\definecolor{Burgundy}{rgb}{.647,.129,.149}
\definecolor{FlickrBlue}{rgb}{0.25,0.60,0.93}
\definecolor{FlickrPink}{rgb}{1.00,0.00,0.52}
\definecolor{FigureBlue}{HTML}{2489BF}
\definecolor{FigureRed}{HTML}{C41C2D}

\definecolor{CI_Grey}{HTML}{808080}
\definecolor{FairGreen}{HTML}{F0F9F0}
\definecolor{ColorHR}{HTML}{52C955} 
\definecolor{ColorAI}{HTML}{008AE4} 
\definecolor{ColorOV}{HTML}{FF3477} 
\definecolor{ColorAO}{HTML}{FF7E54} 
\definecolor{ColorPAO}{HTML}{5F2CC6} 
\definecolor{ColorHR_Light}{HTML}{F0F9F0} 
\definecolor{ColorAI_Light}{HTML}{EBF3FB} 
\definecolor{ColorOV_Light}{HTML}{FDEDF2} 
\definecolor{ColorAO_Light}{HTML}{FFF2ED} 
\definecolor{ColorPAO_Light}{HTML}{EFE9F9} 

\newcommand{\anonymize}[1]{\textsf{ANONYMIZED}}

\newcommand{\ci}[1]{\textcolor{CI_Grey}{$_{\pm #1}$}}  
\newcommand{\shrink}{\vspace*{-.2\baselineskip}}       

\newcommand{\abbr}{Jobindex}
\newcommand{\sset}[1]{\textsf{\emph{#1}}}
\newcolumntype{d}[1]{D{.}{.}{#1}}
\newcommand{\faircell}{\cellcolor{FairGreen}}

\newcommand\fairShadeBox[2][fill=FairGreen]{%
    \tikz[baseline]\node[%
        inner ysep=0pt, 
        inner xsep=2pt, 
        anchor=text, 
        rectangle,
        #1] {\strut#2};%
}
\newcommand\condHRbox[2][fill=ColorHR_Light]{%
    \tikz[baseline]\node[%
        inner ysep=0pt, 
        inner xsep=2pt, 
        anchor=text, 
        rectangle, 
        rounded corners=1mm,
        #1] {\strut#2};%
}
\newcommand{\condHR}{\condHRbox{\textsf{Human recruiting}}}
\newcommand\condAIbox[2][fill=ColorAI_Light]{%
    \tikz[baseline]\node[%
        inner ysep=0pt, 
        inner xsep=2pt, 
        anchor=text, 
        rectangle, 
        rounded corners=1mm,
        #1] {\strut#2};%
}

\newcommand{\condAI}{\condAIbox{\textsf{AI recruiting}}}
\newcommand\condOVbox[2][fill=ColorOV_Light]{%
    \tikz[baseline]\node[%
        inner ysep=0pt, 
        inner xsep=2pt, 
        anchor=text, 
        rectangle, 
        rounded corners=1mm,
        #1] {\strut#2};%
}
\newcommand{\condOV}{\condOVbox{\textsf{Human + AI recruiting}}}
\newcommand\condAObox[2][fill=ColorAO_Light]{%
    \tikz[baseline]\node[%
        inner ysep=0pt, 
        inner xsep=2pt, 
        anchor=text, 
        rectangle, 
        rounded corners=1mm,
        #1] {\strut#2};%
}
\newcommand{\condAO}{\condAObox{\textsf{AI oversight}}}
\newcommand\condPAObox[2][fill=ColorPAO_Light]{%
    \tikz[baseline]\node[%
        inner ysep=0pt, 
        inner xsep=2pt, 
        anchor=text, 
        rectangle, 
        rounded corners=1mm,
        #1] {\strut#2};%
}
\newcommand{\condPAO}{\condPAObox{\textsf{Post-AI oversight}}}

\setcopyright{none}

\begin{document}
\title{Human, Algorithm, or Both? Gender Bias in Human-Augmented Recruiting}
\titlenote{Accepted at the 2026 ACM Conference on Fairness, Accountability, and Transparency (FAccT).}

\author{Mesut Kaya}
\orcid{0000-0003-2305-6683}
\email{mesk@jobindex.dk}
\affiliation{%
  \institution{Jobindex A/S}
  \city{Copenhagen}
  \country{Denmark}
}
\affiliation{%
  \institution{IT University of Copenhagen}
  \city{Copenhagen}
  \country{Denmark}
}

\author{Toine Bogers}
\orcid{0000-0003-0716-676X}
\affiliation{%
  \institution{IT University of Copenhagen}
  \city{Copenhagen} 
  \country{Denmark}
}
\email{tobo@itu.dk}

\renewcommand{\shortauthors}{Kaya and Bogers}


\begin{abstract} 
Recent years have seen rapid growth in the market for HR technology and AI-driven HR solutions in particular. This popularity has also resulted in increased attention to the negative aspects of using AI to support hiring practices, such as the risk of reinforcing existing biases against vulnerable groups based on gender or other sensitive attributes. 
Combining human experience with AI efficiency in making recruiting and selection decisions has the potential to help mitigate these biases, but despite a considerable amount of research on fairness in algorithmic hiring, actual empirical evaluations comparing the fairness of human, AI, and human-augmented decision-making remain scarce. 
In this study, we address this gap by presenting a quantitative analysis of gender bias across three scenarios of a real-world recruitment platform: (1) recruiters searching a CV database manually for relevant candidates, (2) AI-driven matching between candidates and jobs, and (3) a combination of human and AI-driven recruiting. 
We find that human recruiters produce lists of candidates that are fairer in terms of gender than the AI-only solution, with more deliberation by humans resulting in fairer outcomes. However, the combination of human and AI-driven is more than the sum of its parts and produces the fairest candidate lists: interacting with the slate of recommended candidates first before manually searching for additional candidates has a beneficial effect on the gender fairness of the set of candidates that are viewed, clicked, and contacted afterwards. 
Our work provides one of the first empirical comparisons of fairness across human, AI, and hybrid recruiting processes, offering evidence to inform the development of more equitable hiring practices and highlighting the importance of human oversight for mitigating bias in algorithmic hiring.
\end{abstract}

\ccsdesc[500]{Information systems~Recommender systems}
\ccsdesc[500]{Human-centered computing~Empirical studies in HCI}
\ccsdesc[500]{Information systems~Environment-specific retrieval}
\keywords{Fairness, gender bias, human augmentation, recruitment, HR}

\settopmatter{printfolios=true}

\maketitle

\section{Introduction}
\label{sec:intro}






The global market for HR technology is expected to expand from 43.7 billion USD in 2025 to 81.8 billion by 2032 \cite{FBI:2025:HRTech}. 
AI-driven HR solutions are increasingly being used across a wide range of tasks: from helping recruiters identify suitable candidates and assisting job seekers in finding and applying for relevant jobs \cite{Zheng2024SIGIR}, to supporting copywriters in writing job descriptions \cite{wiles2025generative} and enabling automated screening of candidates through AI-scored video interviews that assess their suitability for the position \cite{Kochling:2021:fairness-video-interviews, Hilliard:2022:perception-video-interviews, Kim:2022:AI-video-interviewing}.
The use of AI to support applicant tracking---systems that help automate storing, tracking and searching for applicants---is particularly widespread, with 99\% of Fortune 500 companies reportedly using algorithms and AI to support their hiring practices \cite{Purcell:2025:ATS-usage}.

However, despite promises of increased efficiency, using AI to support and automate hiring processes is not without risks \cite{sanchez2020does}. While AI-driven solutions were initially touted as objective decision-making tools---and therefore intrinsically (more) fair---research has shown that uncritical and unsupervised use of AI risks amplifying rather than reducing existing biases \cite{Fabris2025FairnessHR, Dastin:2022:Amazon-recruiting-tool}. This awareness of the risks of automation has in turn resulted in an increased focus on regulating AI-driven solutions \cite{Sloane2025}.

At the same time, there is an increased realization that the factors and stakeholders that give rise to these biases in algorithmic hiring are complex and multi-faceted in their origins \cite{Sloane2025}. 
In addition to biases in human \cite{steinpreis1999impact, bertrand2004emily} and algorithmic decision making, other sources of bias originate from the structural inequalities in society or stereotyping embedded in cultural norms, social practices, and institutions. These could, for instance, result in gender biases in the world distribution of candidates in different professions \cite{Peng:2019:bias-stages}. 
Biases in communication \cite{Ali:2019:facebook-job-ads, Nagaraj:2023:discrimination, Gaucher:2011:gendered-wording} against vulnerable groups (e.g., women, ethnic minorities, the elderly) have also been well-documented.
%

Given the variety in potential biases, mitigating their influence is therefore likely to require a multi-dimensional approach. A promising example of such an approach is the combination of human experience with AI efficiency in making recruitment decisions. 
Also known as human augmentation, human-in-the-loop, or human oversight \cite{Methnani:2021:human-control}, combining the strengths of both human and AI recruiting could potentially also mitigate their weaknesses.

Some work exists on the {\em perceived} fairness of human and AI-driven decision-making in a recruiting context, with AI being perceived as more fair in some situations \cite{Choung16112024, Qin02102023}.
However, to the best of our knowledge there is no work that actually attempts to {\em quantify} and {\em compare} the fairness of human, AI-driven and hybrid decision-making approaches in a recruiting context.
In this paper, we provide such a quantitative comparison of fairness in different algorithmic hiring scenarios at Denmark's largest job portal. More precisely, we study the degree of {\em gender} bias in {\em candidate recommendation}---recommending relevant candidates for an open vacancy---by recruiters in three scenarios: (i) recruiters searching a CV database {\em manually} for relevant candidates, (ii) {\em AI-driven} matching between candidates and jobs, and (iii) a {\em combination} of human and AI-driven recruiting. 
This leads us to 
the following questions:

\begin{description}

  \item[RQ1] How fair (in terms of gender) is the set of candidates that human recruiters interact with, without the help of AI?

  \item[RQ2] How fair (in terms of gender) is the set of recommended candidates suggested by AI?
  
  \item[RQ3] How fair (in terms of gender) is the set of candidates that human recruiters interact with when supported by AI?
  
\end{description}

We conduct our experimental comparison at \abbr{}, Denmark's largest job portal and recruitment agency. Traditionally, given an open job position, \abbr{}'s recruiters will search for relevant candidates in a database of over 180,000 CVs. In 2023, a recommender system was implemented that shows recruiters a list of recommended candidates before they themselves continue searching for relevant candidates. Engaging with the slate of recommended candidates is optional, making for a clear comparison between these human, AI and hybrid scenarios, providing a useful case study for comparing gender fairness in different scenarios.

\shrink 

\section{Related work}
\label{sec:background}

\subsection{Human Oversight}

\citet{Methnani:2021:human-control} define three different types of human oversight scenarios: (1) human-in-the-loop (HITL),  (2) human-on-the-loop, and (3) human-out-of-the-loop. In a {\em human-in-the-loop} scenario, the human plays an integral role throughout the entire operation, influencing every decision cycle of the system. This is common in high-risk scenarios, such as health informatics \cite{holzinger2016interactive}. 
In a {\em human-on-the-loop}  scenario, the human steps back during operation to a supervisory role, influencing the system by monitoring its behaviours and interfering only as needed, such as in industrial robotics \cite{chen2014human}. 
Finally, in some situations humans lack the expertise, local knowledge, or reaction time to optimally respond to the environment, a {\em human-out-of-the-loop} approach is most appropriate, pushing the human out of the control loop entirely. An example of this are collision detection systems 
\cite{castelfranchi2003automaticity}.
At \abbr{}, recruiters have the end responsibility for the set of contacted candidates and they play a role in the entire process from analyzing the job ad to assessing the recommendations and supplementing them with manual searches. As a result, this workflow is a good example of human-in-the-loop. 
This is also in line with the necessary preconditions for effective human oversight as identified by \citet{Sterz2024oversight}.

There have been several studies of the use of AI-based augmentation in the recruiting domain. \citet{DellAcqua:2022} explored the trade-off between the quality of the AI used vs.\ the potentially adverse impact on human effort. They found evidence for the `falling asleep at the wheel' phenomena in AI-assisted recruiting as more experienced recruiters benefited from lower-quality AI and performed worse than the inexperienced recruiters when receiving higher-quality AI. 
\citet{Lacroux:2022} explored the reactions of recruiters when they are offered AI-based recommendations. 
They found that recruiters exhibited a higher level of trust toward human expert recommendations compared with AI recommendations. In addition, they found that specific personality traits (e.g., extraversion, neuroticism, and self-confidence) have an influence on people's perception and acceptance of the AI recommendations. 

\subsection{Fairness in Human Hiring}

Fairness issues in human hiring and recruitment have been well documented, with unfairness or bias against vulnerable groups based on gender, age, ethnicity, religion, and immigration status. Multiple studies have shown that otherwise identical CVs receive systematically different evaluations by human recruiters depending on the demographic attributes. Applicants with names commonly associated with males are preferred over those associated with females \cite{steinpreis1999impact}, applicants with ‘white’-sounding names receive more callbacks than those with African-American names \cite{bertrand2004emily}, and older applicants face significant discrimination in selection processes \cite{batinovic2023ageism}.
Fairness issues have also been observed in real-world recruitment platforms:  researchers have observed that individuals from immigrant and minority backgrounds receive 4–19\% fewer contact messages 
than other citizens \cite{Hangartner2021}. Disparities also emerge across different types of occupations, with women being disadvantaged in male-dominated fields---a bias that is amplified among older recruiters and firm owners \cite{minssen2025recruiters}. Other studies reveal discrimination against applicants based on religion \cite{di2021muslim} and gendered ethnic discrimination, where men with foreign-sounding names experience greater penalties than women with foreign-sounding names \cite{erlandsson2024gendered}.
Beyond traditional hiring, recent work shows that biased job recommendations generated by LLMs can further alter and limit human decision-making \cite{wilson2025no}. Finally, evidence suggests that gender bias (favoring men) in hiring reflects both stereotypes and prejudices and that it that decreases when women are more qualified but increases when they have children \cite{Gonzalez2019genderstereotypes}.

\subsection{Fairness in Algorithmic Hiring}
Fairness in algorithmic hiring has been widely studied \cite{Fabris2025FairnessHR,chen2023ethics,raghavan2020,sanchez2020does} and research has shown that algorithmic bias can result in discriminatory hiring practices against vulnerable groups based on gender, race, and color \cite{chen2023ethics}. Yet despite a considerable amount of research examining fairness in algorithmic hiring, empirical evaluations comparing the fairness of human, AI, and human-augmented decision-making remain scarce.
%
Some work exists on the perceived fairness of human and AI-driven decision-making in a recruiting context. Employees perceive AI systems used for performance evaluation as both fairer and more accurate than the average human manager \cite{Qin02102023}. In the hiring domain, \citet{Choung16112024} investigated perceptions of AI versus human decision-making in job applications, drawing on evaluations from MTurk participants assessing the fairness of hiring outcomes. They also find that AI is perceived to be fairer than humans. In contrast, \citet{cai2024impact} studies resume screening and reports that applicants tend to perceive AI-based screening as less fair than human decision-making. 
%
Other work has examined the fairness of human-in-the-loop (HITL) systems in the recruiting context \cite{harris2024combining,Peng:2019:bias-stages}. It has been observed that combining HITL and AI tools provides an effective approach to reducing the disparate impact on different age groups in job interview selection \cite{harris2024combining}. One study investigated a HITL system in which AI-generated candidate slates are balanced for gender and then assessed by human decision makers. The results show that while gender bias can be mitigated in some professions, it persists in others where strong human preferences remain \cite{Peng:2019:bias-stages}.
%
The novelty of our work lies in the empirical evaluation of gender fairness in three different recruitment scenarios: (1) human-only matching of CVs to jobs, (2) AI-only matching, and (3) a combination of human and AI-driven matching. 
This provides a unique opportunity to compare these three scenarios in terms of gender fairness.

\section{Methodology}
\label{sec:methodology}

\begin{figure*}[ht!]
  \centering
  \includegraphics[width=0.8\linewidth]{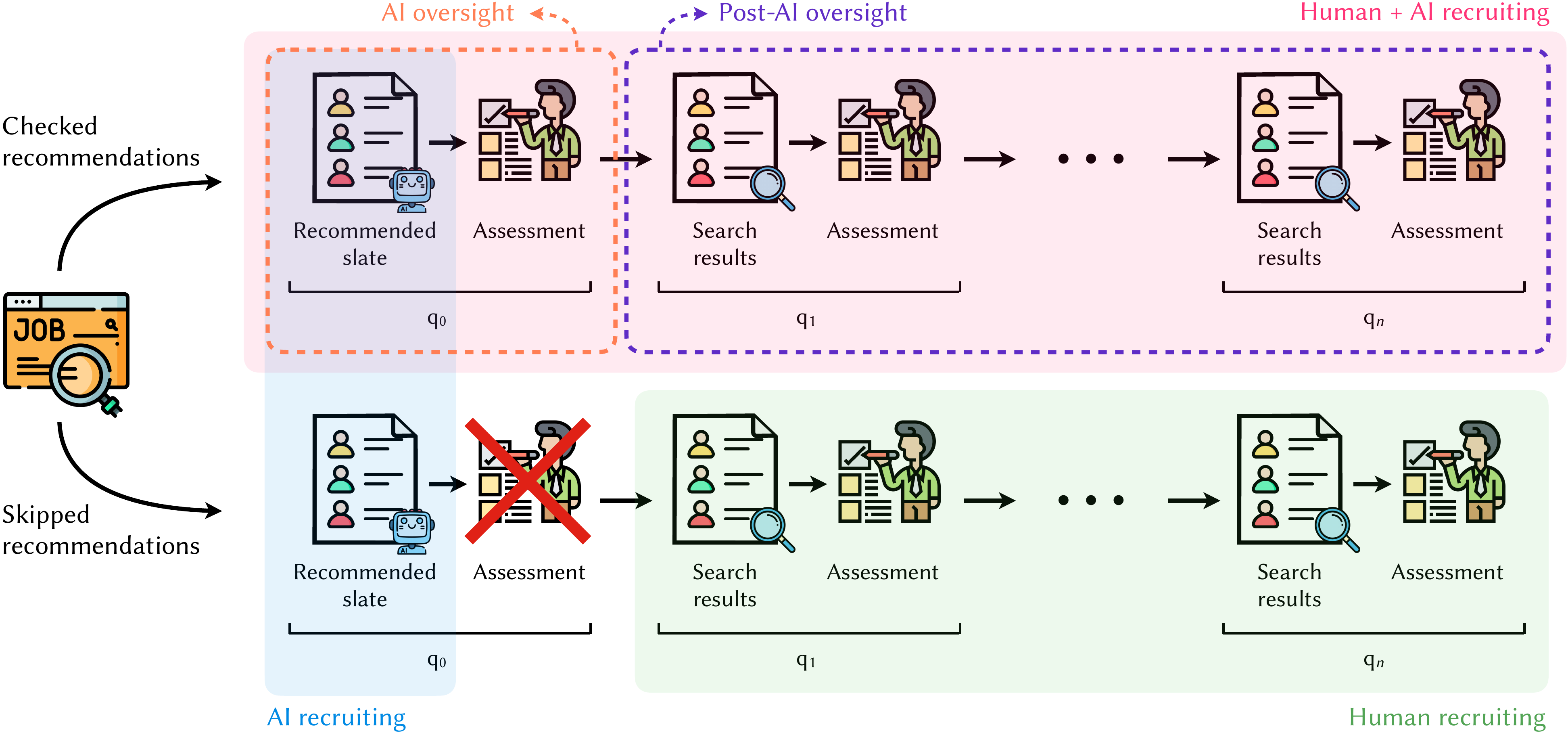}
  \caption{A visualization of the recruiter workflow in the different conditions. Recruiters are first presented with the slate of recommended candidates (\condAI{}). They can choose to skip these recommendations, after which their slate of selected candidates will solely be based on their own searches (\condHR{}). If recruiters decide to engage with the slate of recommended candidates, their workflow is a combination of AI-based recommendation and manual searching (\condOV{}), which we further divide into the AI oversight phase (\condAO{}) and the manual searching phase after having interacted with the recommendations (\condPAO{}).}
  \label{fig:experimental-comparison}
\end{figure*}

\subsection{Experimental design}
\label{sec:experimental-design}

We conducted our experimental comparison of the influence of human and AI involvement in the candidate recommendation process on fairness at \abbr{}, Denmark's largest job portal and recruitment agency. Traditionally, given an open job position, \abbr{}'s recruiters will search for relevant candidates in a database of over 180,000 CVs. 
Access to this database is provided through a standard Solr instance with no dense retrieval or AI-based re-ranking features. The search user interface allows recruiters to search using a combination of free-text keywords, location-based queries, job titles, job categories, and a variety of filters, such as education level, work and management experience, salary expectations, language skills, and contract type (e.g., full-time, part-time). 
Search results show professional and career-related information of the job seekers, such as their current role, job preferences, education, work experience, skills and keywords, location (if provided), and employment conditions. Personally identifiable attributes such as photos, the candidate’s name, date of birth, and gender are masked on purpose.

In 2023, a recommender system was implemented that shows recruiters a list of recommended candidates before they themselves continue searching for relevant candidates. While we are aware that modern search engines can also incorporate more AI-based features, this is not the case at \abbr{}. When we refer to AI in the context of our study, we refer solely to this candidate recommender system. 

Engaging with the slate of recommended candidates is optional, providing an opportunity for a clear comparison between these human, AI and hybrid scenarios in terms of gender fairness. 
\mbox{Figure~\ref{fig:experimental-comparison}} illustrates the AI-assisted recruiting workflow. 
After a recruiter selects a job from the queue, they are presented with the slate of candidates generated by a candidate recommendation algorithm\footnote{This recommendation algorithm is based on a cross-encoder architecture that takes textual input from a job posting and a CV to predict a matching score between the two. Later, these scores are used to recommend the top $N$ relevant CVs for each job posting. This algorithm was chosen to be integrated into the company's production workflow based on multiple iterations of offline and A/B testing \cite{kaya2023spbert}.}. 
%
Choosing to assess (part of) this recommended slate  results in a recruitment workflow that is a combination of AI-based recommendations and manual searching. We refer to this condition as \condOV{}, which is illustrated as the top path in \mbox{Figure~\ref{fig:experimental-comparison}} following ``Checked recommendations''. 
This path is further divided into two stages: the \condAO{} stage, where we look at the fairness of the assessments by human recruiters of the AI-based recommendations, and the \condPAO{} stage after having interacted with the recommendations, where recruiters supplement their slate of selected candidates through manual searching.
%
Recruiters can also choose to ignore the recommendations entirely, after which they immediately start searching for relevant candidates manually. We refer to this condition as \condHR{}, which is illustrated as the bottom path in \mbox{Figure~\ref{fig:experimental-comparison}} following ``Skipped recommendations''.
%
Finally, the fairness of \condAI{}---shaded in light blue in \mbox{Figure~\ref{fig:experimental-comparison}}---refers to fairness of the slate of recommended candidates on its own, independent of human oversight. Here, we combine all recommended slates from checked and skipped jobs, as engagement with the recommendations does not matter here.

Our experimental design shown in \mbox{Figure~\ref{fig:experimental-comparison}} corresponds to a quasi-experimental design because jobs are not assigned to either \condOV{} or \condHR{} in \mbox{Figure~\ref{fig:experimental-comparison}}. Instead, recruiters decide per job whether they will consult the recommendations, often based on their perception of, among other things, the difficulty and industry sector of a job.

\abbr{} does not know whether the contacted candidates are hired by the companies in questions. However, contacted candidates are able to provide positive or negative feedback on the job recommendation from the recruiter, allowing \abbr{} to know how well their recruiters perform. It is also this feedback that was used to train the recommendation algorithm.

\subsection{Data}
\label{subsec:3:data}

The dataset used in our analysis was collected over a 27-month period from April 2023 to July 2025. During this period, \abbr{} recruiters completed 66,654 jobs where they were presented with candidate recommendations at the start of their session. Each job is associated with a single recruiter. A subset of jobs (11.37 \%) has multiple recruiters associated with them. We removed these jobs, because we investigate the effect of individual recruiters on the results. The bottom row in Table \ref{tab:overalldpresults} shows how many jobs were analyzed in the different conditions and stages.
A session consists of at least one query with the initial query showing the recommendation list. After being shown this list, recruiters can start conducting their own searches by issuing a free-text query and activating one or more filters. Recruiters can shortlist any candidate in any of the results lists and contact them immediately or---more commonly---contact the whole list at the end of their session. Recruiters are not forced to select candidates from the list of recommendations.
Each query has an associated set of events related to the the search formulation and source selection/interaction actions of the recruiter. In our analysis we only consider three such events: whether a CV is (i) viewed, (ii) clicked, (iii) or contacted. 
For each job, contacted candidates can respond positively, negatively or not at all. This response data is also logged for each job. 

Across these 58,765 jobs, a total of 1,348,916 candidates were contacted. Of these candidates, 13.67\% responded negatively, 14.86\% responded positively, and the remaining 71.47\% did not respond. On average, recruiters view 105.7 CVs (Md = 84), click 23.7 CVs (Md = 16), and contact 22.95 CVs (Md = 21) per job. 
    
\subsection{Fairness Metrics}
Many different fairness definitions and metrics have been proposed over the last decade to aid in the measurement and assessment of fairness \cite{castelnovo2022clarification}. Each of these metrics formalize different perspectives on how to measure fairness, depending on the specific context, the domain and the problem formulation.
Because of the variety in perspectives on fairness, it is important to select or design fairness metrics that align with the needs of as many stakeholders as possible \cite{Abdollahpouri:2019:fairness, patro2023algorithmic}. 
Our study takes place in the context of algorithmic hiring and we therefore follow the recommendations made by Kaya and Bogers \cite{Kaya2025Mapping}. They interviewed multiple stakeholders in the context of algorithmic hiring---job seekers, recruiters and company representatives---and mapped their findings to fairness metrics that were in line with insights from these stakeholders. 
One of the metrics they marked as relevant is {\em Conditional Demographic (Dis)Parity} (CDP), which is a variant of commonly used group fairness metric Demographic Parity (DP). In our work, we use CDP to measure gender fairness, in line with 
\citet{Kaya2025Mapping}.

\subsubsection{Notation}
We denote $A$ as the categorical random variable representing the protected or sensitive attribute of job seekers---gender identity in our scenario. All other non-sensitive or unprotected attributes are denoted as $X$. A decision-maker $f$ makes a binary decision and provides a `yes' or `no' decision using $A$ and $X$. Formally, $\hat{Y} = f(X,A), \hat{Y} \in \{0,1\}$. For instance, for the human-augmented recruiting scenario explained in \mbox{Figure~\ref{fig:experimental-comparison}}, the candidate recommendation algorithm that generates the slate of recommended candidates is a decision-maker that either recommends a candidate or not. Similarly, a human recruiter is a decision-maker that either chooses to contact a candidate for a job posting or not).  We denote the set of CVs in the CV database as $\mathcal{C}$ (overall population) and the set of qualified candidates as $Q \subseteq \mathcal{C}$.

\subsubsection{Demographic Parity} Demographic Parity (DP) aims to measure the equal acceptance rate across any groups. For example, in the recruiting scenario featured in our study, the decision-making (e.g., recommending, contacting) must be independent of the candidate's gender  ($\hat{Y} \perp\!\!\!\perp A$), which can also be expressed as 
$P(\hat{Y} = 1 \mid A = a) = P(\hat{Y} = 1 \mid A = b), \  \forall a, b \in \mathcal{A}$ \cite{castelnovo2022clarification}.

To compare a protected group $a$ to unprotected group $b$, the DP ratio can be used: 

\begin{equation}
    \mathit{DP}_{a/b} = \frac{P(\hat{Y} = 1 \mid A = a)}{P(\hat{Y} = 1 \mid A = b)}
\end{equation}

In deciding what are appropriate values for the DP ratio, it is common to refer to the `80\% rule' when reporting the DP (or CDP) ratio: while a DP ratio below 80\% is considered unacceptable, DP values above 80\% may be acceptable. This rule is based on the guidelines by the US Equal Employment Opportunity Commission (EEOC) \cite{EEOC1979UniformGuidelinesQA}. It is important to note that the 80\% rule should only be treated as a rule-of-thumb. Because it is based on US practice, it may not be applicable in the EU, so what counts as ``acceptable'' is context-specific. Later, when we present our fairness results in \mbox{Section~\ref{sec:measuring-fairness}}, we use the 80\% for illustration purposes and not to make any normative statement.

\subsubsection{Conditional Demographic (Dis)Parity} 
The Conditional Demographic (Dis)Parity (CDP) metric is a variant of DP that aims to measure the equal acceptance rate across groups in any strata \cite{Watcher2021fairness,castelnovo2022clarification, Kaya2025Mapping}.
Achieving fairness in a human-augmented recruiting setting requires that selection decisions be conditional on candidate qualifications while remaining independent of gender identity.

CDP formally can be defined as $\hat{Y} \perp\!\!\!\perp A \mid Q$. This can also be expressed as $P(\hat{Y} = 1 \mid A = a, Q) = P(\hat{Y} = 1 \mid A = b, Q), \ \forall a, b \in \mathcal{A}$ \cite{castelnovo2022clarification}.
Finally, to compare the protected group $a$ to unprotected group $b$, we can compute CDP ratio ($CDP_{a/b}$) 
similar to DP ratio 
earlier.
For the CDP ratio, $\mathit{CDP}_{a/b} = 1$ indicates perfect parity. Values $\mathit{CDP}_{a/b} < 1$ mean candidates from protected group $a$ are underrepresented, while values $\mathit{CDP}_{a/b} > 1$ mean they are overrepresented. 

When $Q$ corresponds to the set of all CVs in the database ($Q = \mathcal{C}$), CDP is equivalent to demographic parity (DP). While DP can be used to assess systematic fairness between protected and unprotected groups of job seekers across the entire population---in this case the full CV database---CDP instead allows fairness to be assessed for specific subsets of candidates. We note which subsets of candidates are considered in $Q$ where necessary in \mbox{Section \ref{sec:measuring-fairness}}.
\section{Inferring gender}
\label{sec:inferring}

\begin{table}[t!]
    \centering
    \begin{tabular}{lccc}
    \toprule
    & {\bf Female} & {\bf Male} & {\bf Unknown} \\
    \midrule
         Inferred & 16.6\% & 19.6\% & \multirow{2}{*}{22.6\%}  \\
         Self-reported & 20.3\% & 20.8\% & \\

         \bottomrule
    \end{tabular}
    \caption{Descriptive statistics for gender occurrence and inference in the CV dataset.}
    \label{tab:genderdescriptivestats}
\end{table}

In September 2021, \abbr{} decided to cease asking new job seekers to self-report their age and gender identity when adding their CV to the \abbr{} CV database. The reason for this was that they wanted to prevent recruiters from taking these sensitive attributes into account when making relevance assessments about candidates. At the same, photos and names were also removed from the results snippets. Job seekers who registered before September 2021 were able (but not required) to select their gender identity from a restricted set: {\em Male}, {\em Female} or {\em Not provided}. We acknowledge the reductive and potentially harmful nature of this restricted set of options for gender self-identification that was historically adopted by \abbr{} \cite{HafnerGender2025}. For all users who registered their CV after September 2021, their gender was recorded as {\em Not provided}. 

This partial availability of sensitive demographic data is not unique to \abbr{}. \citet{Andrus2021FaacT} has documented the wide variety in demographic data availability and the tensions this can bring \cite{bogen2020awareness}, ranging from not having access to personal data of any kind to being legally required to collect and use demographic data for discrimination assessments.
When sensitive attributes are unavailable and have to be inferred in order to measure fairness, one of the most common approaches is to infer from using proxies \cite{ImanaAuditing2025}, with the first name of the user being the most common \cite{liu2013s,Ghosh2021Inference}. LinkedIn also infers gender using the user's first name and their pronouns\footnote{\url{https://www.linkedin.com/help/linkedin/answer/a517610}}, the latter of which \abbr{} does not record. \citet{bogen2024navigating} provides a more comprehensive overview of the different approaches to inferring gender. 

In line with much of the related work, we also use the job seeker's first name to infer their gender and extract this information from \abbr{}'s CV database. 
Using $\sim$600K self-reported name-gender pairs, we generated a name-gender frequency table and assigned the label {\em Male} or {\em Female} to each job seeker for which the gender was not provided, based on the most frequent label associated with their first name. 
We are aware of more advanced probabilistic techniques for inferring sensitive attributes---such as including geographic information to infer race---but we judged these to not be beneficial when only inferring gender.
To test the performance of this straightforward frequency-based approach, we use stratified 10-fold cross validation. 
Our classifier abstains from assigning a label in some cases:, either due to the first name not being among the first names in our training set or because it was left blank by the user. We report $F_{1}$ score only for the cases for which a definitive prediction could be made, and we additionally report coverage: the fraction of samples receiving a non-unknown prediction by the model. The mean $F_{1}$-score of the model is 99.25 with 94.5\% coverage of samples, which means that for 5.5\% of the samples the model cannot predict {\em Male} or {\em Female}\footnote{We also experimented with a threshold-based approach instead of directly assigning the most frequent label, but observed no performance differences across different threshold values. We also compared per-class $F_{1}$-score values. Although a paired permutation test indicates statistically significant difference in mean $F_{1}$-scores between the {\em Male} and {\em Female} labels (p = 0.0021), the observed difference ($\Delta$$F_{1}$ = 0.0004) corresponds to less than 0.05\% relative difference. This suggests that our classifier results in balanced and robust performance across both the {\em Male} and {\em Female} labels.}.  

%
Table \ref{tab:genderdescriptivestats} presents descriptive statistics for the gender attributes of CVs, including both self-reported and inferred values.

\section{Measuring fairness}
\label{sec:measuring-fairness}

\begin{figure*}[t!]
    \centering
    \begin{subfigure}[b]{0.475\textwidth}
        \centering
        \includegraphics[width=\textwidth]{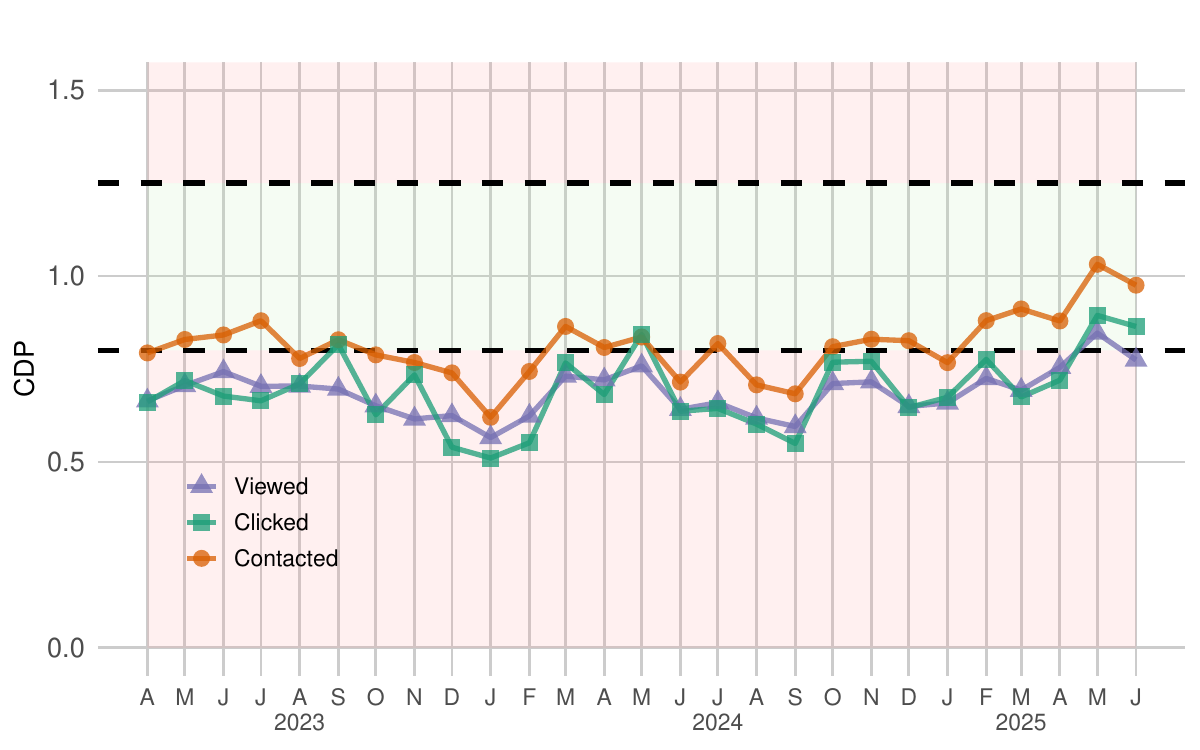}
        \caption{\condHR{}}
    \end{subfigure}
    ~
    \begin{subfigure}[b]{0.475\textwidth}
        \centering
        \includegraphics[width=\textwidth]{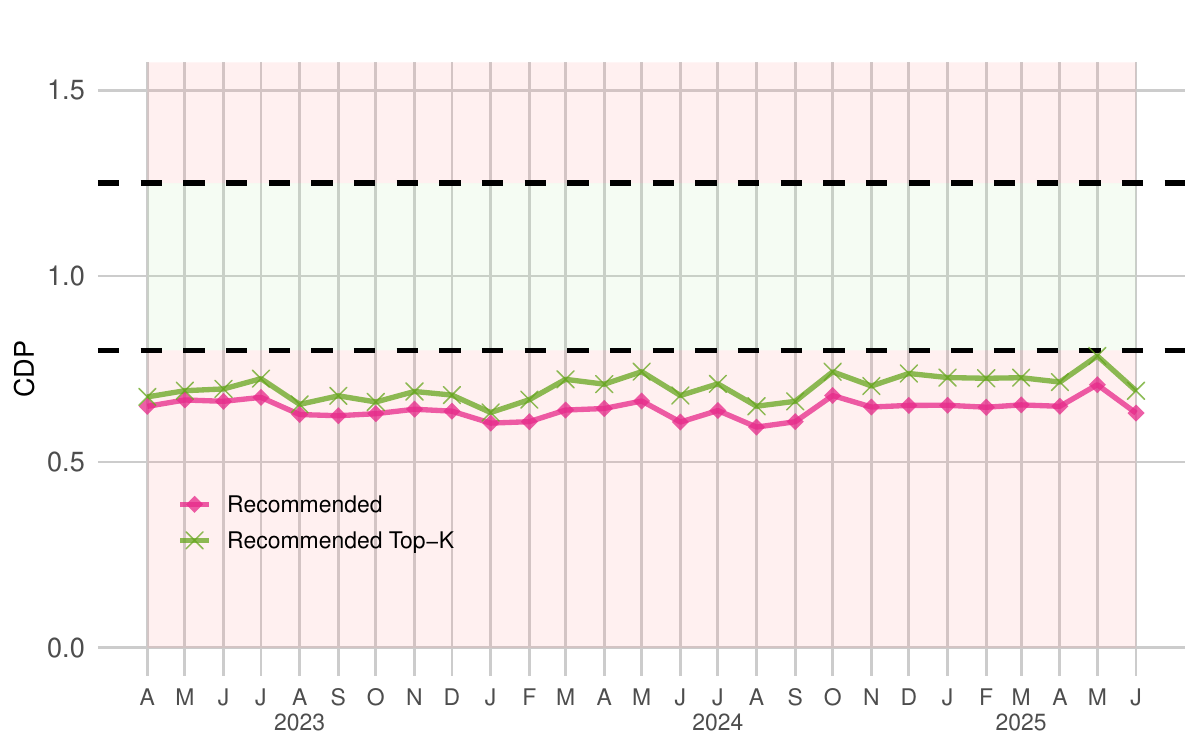}
        \caption{\condAI{}}
    \end{subfigure}\\
    \begin{subfigure}[b]{0.475\textwidth}
        \centering
        \includegraphics[width=\textwidth]{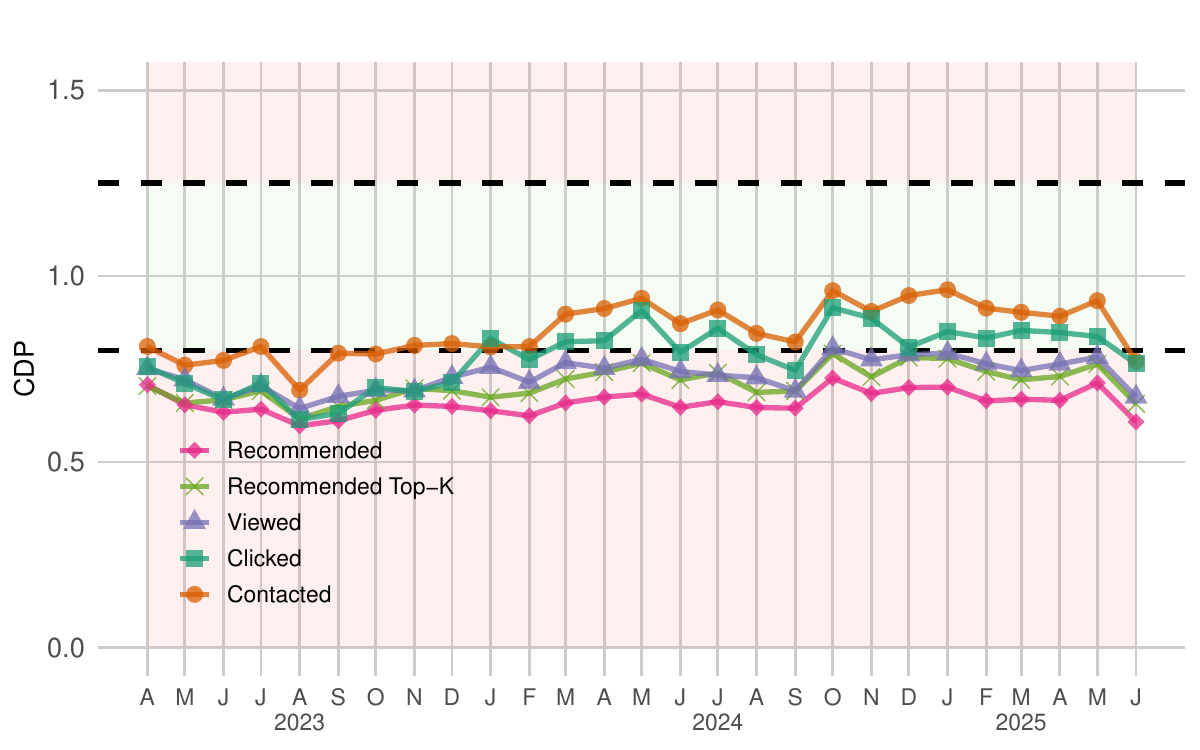}
        \caption{\condOV{}}
    \end{subfigure} 
    ~ 
    \begin{subfigure}[b]{0.475\textwidth}
        \centering
        \includegraphics[width=\textwidth]{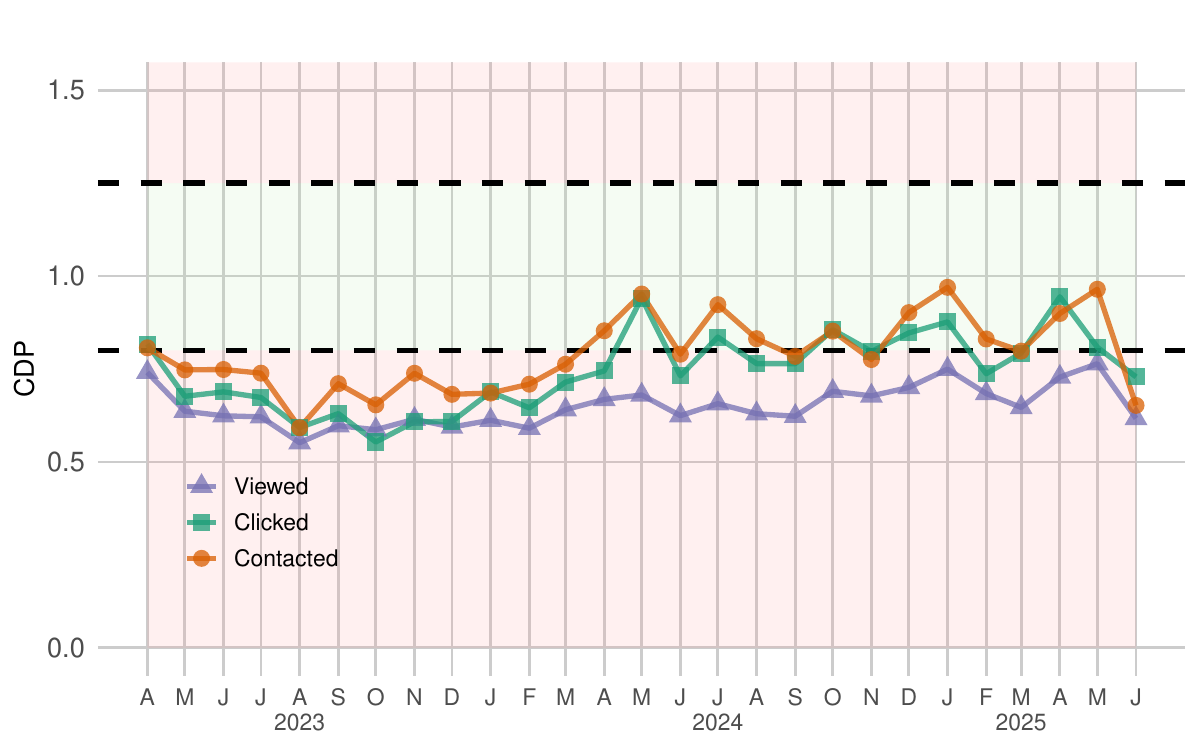}
        \caption{\condAO{}}
    \end{subfigure} \\
    \begin{subfigure}[b]{0.475\textwidth}
        \centering
        \includegraphics[width=\textwidth]{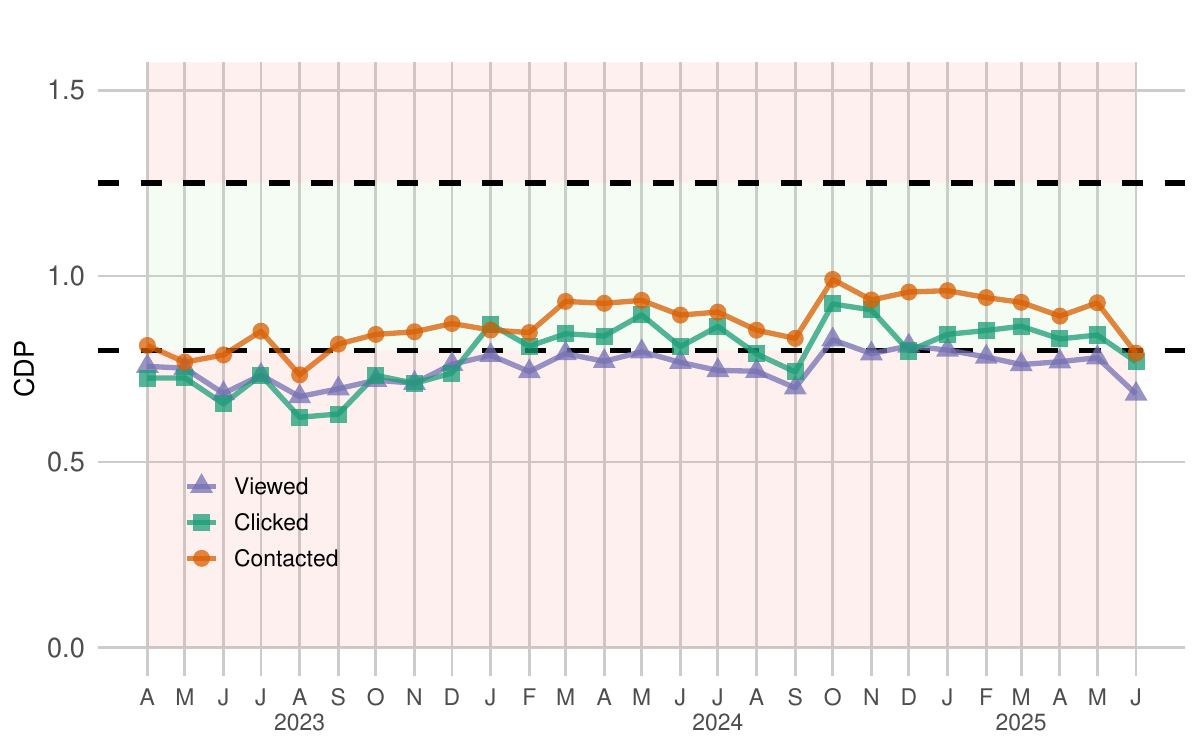}
        \caption{\condPAO{}}
    \end{subfigure}\\
    \caption{These five plots visualize the changes in CDP ratios over time for the different scenarios: (a) \condHR{}, (b) \condAI{}, (c) \condOV{}, (d) \condPAO{}, and (e) \condAO{}. Scenarios with human involvement show three CDP time series for the \sset{Viewed}, \sset{Clicked}, and \sset{Contacted} candidates, while scenarios that involve AI show two CDP time series for the \sset{Recommended} and \sset{Recommended Top-K} candidates. The areas shaded in green in each plot correspond to the 80\% heuristic originating from the EEOC's guidelines, showing a band of 20\% on either side of the perfect CDP score of 1.0, representing perfect fairness. Red-shaded areas indicate CDP ratios outside of these bands.}
    \label{fig:cdp_all}
    \Description[<CDP plots>]{<Details>}
 \end{figure*}

In this section, we present our quantitative comparison of gender fairness in {\em candidate recommendation} in three different scenarios: (1) \condHR{}, where recruiters manually search a CV database for relevant candidates ({\bf RQ1}); (2) \condAI{}, which focuses solely on the candidates suggested by the AI tool ({\bf RQ2}); and (3) \condOV{}, which represents a combination of human and AI-driven recruiting ({\bf RQ3}). In addition, we provide an overview of our dataset in terms of gender skew across job categories in \mbox{Section \ref{sec:category-level-analysis}}. 

In the remainder of this paper, we analyze gender fairness with respect to five different sets of candidates (unless stated otherwise):

\begin{itemize}
  \item {\bf Recommended.} The full set of 100 candidates\footnote{We set size of the recommended candidates list to 100 based on an empirical analysis of how many candidates recruiters view on average. This was determined using interaction log data for a sample of 33,569 jobs that were completed {\em before} our longitudinal analysis started in April 2023. This way, if a recruiter chose to rely solely on the list of recommendations, they would still be able to view a similar number of candidates.} recommended by the AI tool.
    
  \item{\bf Recommended Top-{\em K}.} A subset of the recommended candidates consisting of the top $K$ recommended candidates, where $K$ is the number of candidates actually contacted for each job. Identifying relevant candidates at \abbr{} is not a recall-optimized task; instead, recruiters are instructed to contact between 15 to 25 relevant candidates, depending on different properties of the position, such as complexity and industry. For each job, $K$ is set to the number of contacted candidates for that job. This candidate set represents the hypothetical scenario in which $K$ recommended candidates are automatically contacted by recruiters without intervention. 

  \item{\bf Viewed.} Candidates whose search result snippet has been seen by the recruiters, either in the recommendation list, in the manual search results or both. 
    
  \item{\bf Clicked.} A subset of the viewed candidates that recruiters have clicked on to access more detailed information about the candidate for further assessment. 

  \item{\bf Contacted.} The set of candidates whom recruiters have identified as relevant and contacted about the open job. These have often been clicked on by the recruiter prior to this, but this is not required; shortlisting can also be done based on viewing the search snippet alone.
    
\end{itemize}

For each job posting $j$, we denote these five subsets by using subscript $j$, e.g., $\mathit{Viewed}_{j}$, $\mathit{Contacted}_{j}$, etc. For a month $t$, we take the disjoint union of the candidate sets, for instance $\mathit{Viewed}_t = \biguplus_{j \in J_t} \mathit{Viewed}_j$, where $\biguplus$ denotes disjoint union of sets (as a candidate may appear multiple times). $J_t$ is the set of all jobs that recruiters completed in month $t$. Then, for each month, we computed the {\em CDP} ratio of the aggregate sets separately for \sset{Viewed}, \sset{Clicked}, \sset{Contacted}, \sset{Recommended} and \sset{Recommended Top-K} sets. To measure the monthly {\em CDP} ratio for different sets, the set $Q$ is defined as the set of available and online CVs. The latter is an important distinction because job seekers can take their CV offline and render it not searchable.  

We calculate CDP ratios based on gender, which is a combination of self-reported and inferred gender (as described in \mbox{Section \ref{sec:inferring}}). It is possible that errors in the gender inference process could possibly have an effect on our findings. However, when re-running the same analyses using only the CVs where gender was self-reported, all trends reported below hold, albeit with slightly different scores.

When comparing the difference between different scenarios in terms of statistical significance, we compare the final \sset{Contacted} sets of different scenarios to each other where possible as these represent the actual impact the different scenarios have. Otherwise, we compare the fairness of the human \sset{Contacted} set to the AI-generated \sset{Recommended Top-K} set---representing a hypothetical scenario in which recommended candidates are contacted without human intervention as both constitute the final decision stage that determines candidates. To test whether the mean paired difference between two sets is statistically significant, we conducted permutation tests on paired differences using 100,000 permutations.

\begin{table*}
  \centering
  \small
  \begin{tabular}{l c c c|c c}
    \toprule
    ~ & \condHR{} & \condAI{} & \condOV{} & \condAO{} & \condPAO{} \\
    \midrule
    {\bf Recommended} & - & 0.642 \ci{0.005} & 0.659 \ci{0.006} & - & - \\
    {\bf Recommended Top-{\em K}} & - & 0.699 \ci{0.007} & 0.710 \ci{0.009} & - & - \\
    {\bf Viewed} & 0.687 \ci{0.012} & - & 0.734 \ci{0.008} & 0.650 \ci{0.010} & 0.754 \ci{0.008} \\
    {\bf Clicked} & 0.693 \ci{0.019} & - & 0.783 \ci{0.016} & 0.743 \ci{0.020} & 0.792 \ci{0.016} \\
    {\bf Contacted} & 0.813 \ci{0.017} & - & 0.854 \ci{0.014} & 0.791 \ci{0.019} & 0.876 \ci{0.013} \\
    \midrule
    {\bf Dataset} & {\sc Skip} & {\sc Check} + {\sc Skip} & {\sc Check} & {\sc Check} & {\sc Check} \\
    {\bf No. of jobs} & 21,695 & 58,765 & 37,070 & 37,070 & 37,070  \\
    \bottomrule
    \end{tabular}
    \caption{Mean CDP ratio scores over the 27-month period for all five scenarios. Each scenario also contains the number of jobs in the subset of the data and whether it contains jobs that were skipped, checked or both.}
    \label{tab:overalldpresults}
\end{table*}

\subsection{Human recruiting}
\label{sec:human-only}
To determine how fair the interactions and selections by the human recruiters are when they do not use AI ({\bf RQ1}), we consider only the cases where recruiters skipped the recommendations ({\sc Skip}) and instead identify and contact all candidates based solely on their manual querying (marked as \condHR{} in the bottom right of the \mbox{Figure \ref{fig:experimental-comparison}}). We study fairness on three of the five candidates subsets: (i) \sset{Viewed}, (ii) \sset{Clicked} and (iii) \sset{Contacted}. The other two candidate subsets (\sset{Recommended} and \sset{Recommended Top-K}) are not relevant here as recommendations are ignored in the \condHR{} scenario.

\mbox{Figure \ref{fig:cdp_all}a} shows how fairness changed over time across the three sets of \sset{Viewed}, \sset{Clicked} and \sset{Contacted} candidates. The workload of \abbr{} recruiters varies considerably throughout the year and during busy periods recruiters are instructed to spend less time per job. This is a plausible explanation for the variability in (all of) the time series in \mbox{Figure \ref{fig:cdp_all}}. The three time series are also strongly and significantly correlated with each other with Pearson's $r$ values ranging between 0.79 and 0.89 and all $p$-values below 0.001. Pearson's $r$ correlations between first-differenced series are also strong, ranging between 0.62 and 0.75, and statistically significant with all $p$-values below 0.001. This is understandable as \sset{Contacted} candidates will nearly always be \sset{Viewed} and \sset{Clicked} first as well.

In general, we are more interested in analyzing the overall differences than the month-to-month variations. We observe that female candidates are consistently underrepresented in the \sset{Viewed} and \sset{Clicked} sets, with the \sset{Clicked} set being slightly fairer than the \sset{Viewed} set. In the \sset{Contacted} set, female candidates also tend to be underrepresented, although the disparity is less pronounced than in the other two sets. 
\mbox{Table \ref{tab:overalldpresults}} shows the mean CDP ratios over a 27-month period. The mean CDP scores for \sset{Viewed} and \sset{Clicked} are both below 0.7, while \sset{Contacted} is slightly above 0.8. The CDP score for \sset{Contacted} is significantly higher than for \sset{Clicked} ($\Delta_\mathit{CDP} = 0.120$, $p < .001$) and \sset{Viewed} ($\Delta_\mathit{CDP} = 0.126$, $p < .001$). %

Overall, it appears that more effort spent on interacting with the candidates results in a fairer set of candidates---\sset{Contacted} candidates are more likely to have received the greatest scrutiny, followed by \sset{Clicked} and finally \sset{Viewed} candidates. 
In the last two months of our dataset, we observe an almost perfect gender parity for the set of \sset{Contacted} candidates. This is not due to any training the recruiters have received around this time. While the time series for \sset{Contacted} candidates does show a weak increasing trend according to a modified Mann-Kendall test using Hamed and Rao variance correction \cite{hamed1998modified} (Sen's slope = 0.0033 units/month, $\tau$ = 0.20), this trend was not statistically significant after accounting for serial correlation ($z_c = 1.08$, $p = .28$). The other two sets also show weak increasing trends that are not statistically significant.

\subsection{AI recruiting}
\label{sec:ai-only}
To answer {\bf RQ2}, we study how fair the AI-based candidate recommendations are (shown as \condAI{} in \mbox{Figure \ref{fig:experimental-comparison}}). A slate of 100 recommended candidates was generated for each job, regardless of whether the list was assessed by a recruiter, so here we analyze both {\sc Skip} jobs and jobs for which the recruiters did interact with the recommendations ({\sc Check}). We study fairness on two of the candidate subsets: \sset{Recommended} and \sset{Recommended Top-K}.  

Figure \ref{fig:cdp_all}b shows the change in fairness over time. These two time series are strongly correlated with a Pearson's $r$ of 0.85 ($p < 0.001$), also when correlating the first-differenced series ($r = 0.95$, $p < .001$).
Female candidates were consistently underrepresented in the set of recommended candidates, possibly due to historical biases learned by the recommendation algorithm from the period when recruiters were still shown a candidate's photo and name in the results snippets. 
As shown in \mbox{Table \ref{fig:cdp_all}}, the gender distribution of the candidates in the \sset{Recommended Top-K} appear to be slightly fairer compared to \sset{Recommended} with a mean CDP score of 0.699 vs.\ 0.642 ($\Delta_\mathit{CDP} = 0.057$, $p < .001$), suggesting that gender bias is less pronounced among the higher-ranked recommendations.  
CDP scores for \condAI{} (\sset{Recommended Top-K}) were significantly ($\Delta_\mathit{CDP} = 0.114$, $p < 0.001$) lower than for \condHR{} (\sset{Contacted}), suggesting that relying on fully-automated candidate recommendation risks amplifying gender unfairness as it may systematically underrepresent female candidates.

Both time series show a weak increasing trend, but it was only statistically significant for \sset{Recommended Top-K} according to a modified Mann-Kendall test (Sen's slope = 0.0021 units/month, $\tau = 0.32$, $z_c = 2.29$, $p = .022$). As the recommendation model was never retrained during the 27-month study period, this suggests that this increase could be due to a change in the makeup of the CV database in terms of gender.

\subsection{Human-augmented recruiting}
\label{sec:human-augmented}
When recruiters interact with candidates from the recommendation list, we can study the effects of this combination of AI-based candidate recommendation and manual searching on gender fairness ({\bf RQ3}). We first analyze this combined \condOV{} scenario, followed by a breakdown into the \condPAO{} and \condAO{} phases. %

\subsubsection{Human + AI recruiting}
To investigate gender fairness in \condOV{}, we compare all five candidate subsets for the {\sc Check} jobs. 
\mbox{Figure \ref{fig:cdp_all}}c shows the change in gender CDP ratios over time for all five candidate subsets while \mbox{Table \ref{tab:overalldpresults}} shows the mean CDP scores over the 27-month period. 
All five time series corresponding to the candidate subsets exhibit statistically significant strong correlations. What we also observe in \mbox{Figure \ref{fig:cdp_all}}c is a consistent increase in gender fairness in the following order: \sset{Recommended} $\rightarrow$ \sset{Recommended Top-K} $\rightarrow$ \sset{Viewed} $\rightarrow$ \sset{Clicked} $\rightarrow$ \sset{Contacted}. The \sset{Contacted} subset of candidates has the highest mean CDP score at 0.854, which represents a significantly ($\Delta_\mathit{CDP} = 0.063$, $p < 0.05$) fairer gender distribution than in the \condHR{} scenario at 0.813, suggesting that the combination of human and AI candidate recommendation provides the fairest outcomes. The CDP score of \sset{Contacted} for \condOV{} is significantly higher than in the \condAI{} (\sset{Recommended Top-K}, $p < 0.001$) scenario. 

The progression in CDP score improvement among the five candidate subsets is most likely due to the intensity of the recruiters' engagement. Shortlisting the final set of \sset{Contacted} candidates represents the most deliberate evaluation of candidates of all, followed by \sset{Clicked} and \sset{Viewed} candidates. This increase in scrutiny of the candidates also appears to lead to fairer outcomes. As in \mbox{Section \ref{sec:ai-only}}, the set top-ranked AI-recommended candidates are more fair.

All of this suggests that human oversight indeed has the potential to mitigate biases as the mean CDP ratio drops when recommendations are skipped (\condHR{}) compared to when recruiters interact with them (\condOV{}). What is not clear is in which phase the majority of this bias mitigation occurs: in the human oversight over the AI-provided recommendations (\condAO{}) or in the manual searching that follows (\condPAO{})? We answer this question in Sections \ref{sec:pure-oversight}-\ref{sec:post-ai-oversight}.
Finally, over the course of the 27-month period, gender fairness of all of the five candidate subsets for \condOV{} has increased significantly according to a modified Mann–Kendall test with all $p$-values at least below .05. The only possible explanation for this we can offer is that \abbr{} recruiters needed time to get to know the strengths and weaknesses of the slate of recommended candidates, allowing them to better balance the fairness of their selections over time.

\subsubsection{AI oversight}
\label{sec:pure-oversight}

\condAO{} is one potential stage of bias mitigation, which corresponds solely to recruiters interacting with the slate of recommended candidates ($q_0$ in \mbox{Figure \ref{fig:experimental-comparison}}). 
\mbox{Figure \ref{fig:cdp_all}d} shows that that the fairness of candidates that were \sset{Clicked} and/or \sset{Contacted} is higher than the pure recommendations in \mbox{Figure \ref{fig:cdp_all}b}.
While the gender fairness of the recommendations (\condAO{}) is lower than that of manual searching (\condHR{})---at 0.791 to 0.813---human oversight does improve the fairness of the recommendations on their own (\condAI{}) with a maximum CDP score of 0.699, resulting in more equitable shortlisting. 
This pattern indicates that human oversight has the potential to serve as an effective corrective mechanism to algorithmic bias, though the improvement 
cannot fully eliminate it. Recruiters may be aiming, consciously or subconsciously, for increased diversity in their 
shortlist. 

\subsubsection{Post-AI oversight}
\label{sec:post-ai-oversight}

Perhaps the most likely stage of bias mitigation is the \condPAO{} phase where recruiters search for additional candidates after having interacted with the slate of recommended candidates ($q_1,\ldots,q_n$ in \mbox{Figure \ref{fig:experimental-comparison}}). 
\mbox{Table \ref{tab:overalldpresults}} shows that this stage results in candidate sets that exhibit the highest gender fairness. The subset of \sset{Contacted} candidates found manually has a mean CDP of 0.876, which is significantly higher than both the candidates contacted from the recommendation list at 0.791 (\condAI{}, $\Delta_\mathit{CDP} = 0.085$, $p < 0.001$) and the candidates contacted based solely on manual searches at 0.813 (\condHR{}, $\Delta_\mathit{CDP} = 0.063$, $p<0.01$).
These findings are evidence of two phenomena. First, it shows that bias mitigation by human recruiters takes place both in the recommendation stage and the manual search stage. Second, despite the fact that manual searching results in fairer candidate sets than automatic recommendations, the combination of the two provides the fairest sets of candidates. This suggests that the combination of AI and human recruiting is more than the sum of its parts and that inspiration or influence from inspecting the recommendations has a beneficial effect on the gender fairness of the set of candidates that are \sset{Viewed}, \sset{Clicked}, and \sset{Contacted}.
While manual searching in the \condPAO{} stage may result in the highest CDP score, this is conditional on the recruiters' interaction with the recommended slate beforehand; manual searching alone results in less fair outcomes, despite \condAO{} dragging the CDP scores of \condPAO{} down.

\begin{table*}[ht!]
    \centering
    \small
  \begin{tabular}{p{0.25\linewidth}cc|p{0.35\linewidth} cc}
    \toprule
    {\bf Female} & {\bf Female \%}&{\bf CDP ratio} & {\bf Male} & {\bf Male \%} & {\bf CDP ratio} \\
    \midrule
         Medical secretary&88.3 & 2.43 \ci{0.44} & Plumbing and sheet metal work  & 85.4&0.66 \ci{0.04}\\
         Secretary and reception & 77.6& 1.01 \ci{0.02} & Mechanics and auto & 82.1&0.62 \ci{0.01}\\
         \faircell{}Nurse and midwife & \faircell{}70.5& \faircell{}0.87 \ci{0.02} & Electrician & 81.3&0.59 \ci{0.01} \\
         \faircell{}Textiles and crafts& \faircell{}69.4 & \faircell{}0.92 \ci{0.07} & Iron and metal & 81.2& 0.49 \ci{0.02}\\
         \faircell{}Public administration & \faircell{}66.0&\faircell{}0.88 \ci{0.02} & Construction and civil engineering & 78.5& 0.58 \ci{0.01}\\
         \faircell{}Psychology and psychiatry & \faircell{}65.2&\faircell{}0.92 \ci{0.03} & Carpenter and joiner & 78.3 &0.64 \ci{0.04} \\
         Translation and language & 64.5&  1.13 \ci{0.03} & Property service & 77.7&0.49 \ci{0.02} \\
         \faircell{}Childcare & \faircell{}63.7&\faircell{}0.73 \ci{0.06} & Telecommunications and data communication & 76.2&0.37 \ci{0.08} \\
         Hairdressing and personal care & 63.1& 3.19 \ci{0.52} & \faircell{}Executive management and board of directors & \faircell{}75.3&\faircell{}1.39 \ci{0.06} \\
         Dentist and clinic staff & 63.0&1.87 \ci{0.29} & Wood and furniture industry & 74.2&0.77 \ci{0.08} \\
    \bottomrule
    \end{tabular}
    \caption{Overview of the top-10 gender-dominated job categories for both {\em Male} and {\em Female} contacted candidates, sorted by the share of CVs within each gender category. The mean CDP ratio is included along with the standard error. Categories shaded in \fairShadeBox{green}  are categories where the opposite gender is overrepresented (as evidenced by a CDP ratio lower than resp. higher than 1).}%
    \label{tab:genderdominatedcategories}
\end{table*}

\subsection{Additional analyses}
\label{sec:additional-analyses}

\subsubsection{Job category}
\label{sec:category-level-analysis}

In the labor market, some occupations from certain job categories are dominated by a single gender, which is also known as {\em occupational gender segregation} \cite{froehlich2020gender}. To investigate gender fairness for a range of different occupation categories, we performed an analysis of the gender fairness in the most skewed job categories\footnote{There are 87 job categories. At \abbr{}, copywriters manually assign one or more of these categories to each job posting upon submission. Using these manual annotations and the database of historical interactions between job seekers and jobs, \abbr{} has developed an automatic classifier for assigning job categories to CVs. It automatically assigns at least one of the 87 job categories to each CV.}.
For each category, we compute the monthly CDP ratio of \sset{Contacted} candidates. Specifically, for a given category $c$ and month $t$, we define the set \sset{Contacted}$_{ct}$ as all contacted candidates belonging to category $c$. To compute the CDP ratio, $P(\hat{Y} = 1 \mid A = a, Q)$, the set $Q$ consists of all available CVs in month $t$ that fall under category $c$.
To identify the top 10 male-dominated and top 10 female-dominated categories, we sorted all job categories by the share of male and female CVs assigned to that job respectively.
\mbox{Table \ref{tab:genderdominatedcategories}} shows the top 10 female- and male-dominated job categories with their share of female/male CVs as well as the average CDP ratio. Job categories shaded in green indicate that, on average, the sets of \sset{Contacted} candidates feature an over-representation of the opposite gender. For instance, despite {\em Childcare} being a job category with 63.7\% female CVs, male CVs are still overrepresented in this category.
Among the female-dominated job categories {\em Nurse and midwife}, {\em Textiles and crafts}, {\em Public administration}, {\em Psychology and psychiatry}, and {\em Childcare} are highlighted in green, suggesting that males are relatively over-represented among contacted candidates in these female-dominated fields, implying that recruiter behavior may favor men in those specific contexts. 
However, for male-dominated categories, only {\em Executive management and board of directors} highlighted in green, indicating that females are relatively overrepresented among contacted candidates in this male-dominated domain. In 13 out of 87 categories, female candidates are over-represented (15\%), meaning that in the remaining 85\% of categories there is a gender bias favoring male candidates. 
These results indicate a more nuanced pattern of gender bias: over-representation of male candidates occurs more frequently in the top female-dominated categories than over-representation of female candidates in the top male-dominated categories. A possible explanation for these counterintuitive CDP values could be that recruiters are aware of this gender skew and actively try to (over)compensate for the perceived gender gap.


\subsubsection{Recruiter characteristics}
In order to investigate whether there is an recruiter effect on gender fairness as measured by CDP, we constructed a two-level mixed-effects model with completed jobs (Level 1) nested within recruiters (Level 2). Our predictor variables included recruiter experience, job complexity, and various variables related to gender skew. We found that only 2.7\% of the variance could be attributed to between-recruiter differences. As a result we omit the details of this mixed-effects model.


\subsubsection{Response rates}
While \abbr{} recruiters are rarely made aware of the final hiring decisions, they can see the {\em positive response rate}, the share of \sset{Contacted} candidates that responded positively to the job recommendation. We calculate the Pearson's correlation coefficient between this positive response rate and the CDP ratio for each job and found a negligible but statistically significant positive relation ($r(57,143) = 0.067$, $p < .0001$). This means that 0.45\% of the increase in PRR can be explained by the gender fairness of the candidate set, which in turn suggests that the expected trade-off between fairness and accuracy is weak as an increase in fairness does not result in a decreased positive response rate.

\section{Discussion \& Conclusions}
\label{sec:discussion}
In this paper, we presented the results of quantitative comparison of gender fairness of human recruiting, AI-driven recruiting and the combination of human and AI-driven recruiting. We first investigated the gender fairness of candidate selection by human recruiters by looking at the candidates they \sset{Viewed}, \sset{Clicked} and \sset{Contacted} while manually searching the CV database ({\bf RQ1}). We found that female candidates are consistently under-represented, but that it varies considerably from month to month. This is most likely due to variations in work load, resulting in less time for recruiter to reflection on the makeup of the candidate set they interact with. We also found that as recruiter spend more effort on interacting with the candidates' CVs, the level of gender fairness increased.  

When exploring the gender fairness of the recommendations generated by our candidate recommendation algorithm ({\bf RQ2}), we found that they were less fair than the candidates identified by the human recruiters with female candidates again being under-represented consistently, although candidates recommended at the higher ranks were more fair. The increased unfairness could be due to historical biases from human recruiters being reproduced by the algorithm from when they were still candidate photos, gender and name in the CV result snippets. 

The combination of human and AI-driven recruiting produced the fairest candidate lists of all three scenarios ({\bf RQ3}). This bias mitigation takes place both in the initial interaction with the recommendations and in the manual search stage afterwards as both phases produce more gender-fair candidate lists than when conducted separately. This shows that the combination of human and AI recruiting is more than the some of its parts and that the interaction between them further reduces unfairness. Finally, we observed that gender fairness slowly increased over time, possibly due to a learning effect of working with the AI-driven recommendations.

When digging deeper, we found that gender fairness varies strongly by job category and that female-dominated jobs more often suffer from male over-representation than vice versa. Recruiter differences were not found to have any meaningful impact on gender fairness, possibly due to the standardized, proprietary nature of the CV search engine used by \abbr{} recruiters.

Our work provides one of the first empirical comparisons of fairness across human, AI, and hybrid recruiting processes, offering evidence to inform the development of more equitable hiring practices and highlighting the importance of human oversight for mitigating bias in algorithmic hiring. However, as human and AI-based recruiting can be combined in a myriad of ways, so our study should only be seen as a first step towards better understanding how human-AI collaboration can impact fairness.

Our work also has other limitations. While we can conclude that the combination of human and AI-based recruiting is more than the sum of its parts, we were not able to qualify exactly how this interaction takes place. Surfacing how recruiters were affected subconsciously in their manual searches by interacting with the AI-based suggestions just before is challenging at best.
The effect of job category also needs to be investigated more: we only reported the CDP ratios for male and female-dominated candidates that were contacted by the recruiters, but it is possible that gender fairness varies strongly between scenarios and different candidate sets. More work is needed to understand these effects. The category-specific analysis in \mbox{Section \ref{sec:category-level-analysis}} is a step in the right direction, but a more detailed assessment of fairness relative to job- and category-relevant candidate pools---ideally based on structured qualifications or job-category strata---is left for the future work. 
Finally, our study does not take into account constraints at other HR companies or in other domains, which could impact the generalizability of our findings beyond \abbr{} and the HR domain in general. We also acknowledge that our findings are culture- and country-specific.

One avenue for future work is to study fairness with regard to other sensitive attributes, such as age or ethnicity. Another promising direction is the development of fairness-aware recommendation algorithms, along with systematic and longitudinal analyses of their impact on fairness metrics. 
Finally, we acknowledge that our findings represent only a first step toward understanding fairness in scenarios where human and AI-driven recruiting are combined. Conducting periodic audits to assess and compare the fairness of human, AI-based, and hybrid recruiting processes is an important future task.

\section*{Generative AI Usage Statement}
During the preparation of this work, the author(s) used ChatGPT in order to refine academic language in a handful of situations. After using this tool, the author(s) reviewed and edited the content as needed and take full responsibility for the content of the publication.

\section*{Ethical Considerations Statement}
In this study, self-identified gender was limited to specific categories `Male' and `Female', which risks essentializing complex gender identity into a narrow set of biological or culturally dominant norms by reinforcing the idea that these categories are fixed \cite{HafnerGender2025}. In addition, using candidates' first names as a proxy for inferring gender could cause unequal distribution of misrecognition \cite{lockhart2023name}, for instance for applicants from different cultural backgrounds. Information about candidates’ cultural backgrounds is not available at \abbr{}, so inferring cultural background and accounting for this in bias mitigation was beyond the scope of this submission.

Only the first author employed at the company had access to the dataset and with the approval of the \abbr{}’s DPO, we ran all analysis and experiments.

\begin{acks}
The work by Mesut Kaya and Toine Bogers was supported by the FairMatch project (Innovation Fund Denmark grant number 3195-00003B). The work by Toine Bogers was also supported by the Pioneer Centre for AI (DNRF grant number P1).
\end{acks}

\bibliographystyle{ACM-Reference-Format}
\bibliography{facct2026-preprint}

\end{document}